\documentclass[prb,twocolumn]{revtex4}   
\usepackage{graphicx}

\begin{document}

\title{
Spectral functions in itinerant electron systems with geometrical frustration}

\author{
Yoshiki Imai and Norio Kawakami
}

\affiliation{
Department of Applied Physics, 
Osaka University, Suita, Osaka 565-0871, Japan}

\date{
Submitted, Dec. 22, 2001
}

\begin{abstract}
The Hubbard model with geometrical frustration is 
investigated in a metallic phase close to half-filling. We calculate the 
single particle spectral function for the triangular 
lattice within dynamical cluster approximation, which is further
combined  with non-crossing approximation and fluctuation exchange approximation to treat the resulting cluster Anderson model. It is shown that frustration
due to non-local correlations suppresses short-range antiferromagnetic 
fluctuations and thereby assists the formation of heavy quasi-particles
near half-filling.
\end{abstract}

\pacs{Valid PACS appear here}%

\maketitle
Recently, geometrically frustrated metallic systems have attracted 
much attention, for which a wide variety of interesting phenomena have 
been discovered.  For instance, the compound
${\rm LiV_{2}O_{4}}$ with pyrochlore 
structure, which is given by a corner-sharing network of 
tetrhedra of V ions, exhibits a heavy fermion behavior \cite{Kon97}. 
Also, another pyrochlore compound ${\rm Y(Sc)Mn_{2}}$ \cite{Bal96} 
shows a quantum spin-liquid behavior
down to low temperatures. These interesting phenomena are 
considered to be closely related to geometrical
frustration \cite{Kap00,Iso00,Cap01,Lac01,Fuj01,Kas01} induced by the strong 
Coulomb interaction
with specific geometry of the lattice.  

Among theoretical approaches to strongly correlated electron systems, 
dynamical mean field theory (DMFT) \cite{Geo96} is one of the most 
successful methods to describe various physical properties. 
In DMFT the self energy is given as a local quantity, which is justified 
in the limit of large dimensions \cite{Met89,Mul89}. This method 
has been known as a good approximation even in three dimensions.
However, since the self energy becomes local in DMFT, 
 non-local charge and spin correlations, which are  
essential for treating geometrical frustration,
 cannot be properly described in this framework.
In order to deal with such non-local correlations, 
various approaches beyond
DMFT have been proposed \cite{Don94,Sch95,Tra98,Lic00,Kot01}. 
Among others, dynamical cluster
approximation (DCA) \cite{Het98,Het00,Mai00,Mou01,Jar01}  
may provide a systematic way to incorporate non-local correlations.
In DCA, the lattice problem is replaced by the corresponding 
cluster one embedded in an effective medium determined self-consistently. 
 This method has the following 
nice features: the algorithm is fully causal and the approximation
can be improved systematically
if the cluster size is increased.

In this paper, we investigate the effects of geometrical 
frustration on the Hubbard model. In particular, we focus on the triangular
 lattice, which is known as a prototypical system 
having strong geometrical frustration. Since it is important to incorporate
short-range correlations systematically, we 
employ DCA, which is further combined with non-crossing 
approximation (NCA) \cite{Bic87,Pru89} to solve the local problem. 
We also use a weak-coupling approach by means of
fluctuation exchange approximation (FLEX) \cite{Bic89} for the
local problem, which should be
 complementary to the treatment of NCA.
By calculating the one-particle spectral function as well as  the 
total density of states in these approximations, 
we discuss how geometrical frustration affects the formation of 
heavy quasi-particles in a metallic phase close to half-filling. 

\begin{figure}[htb]
\includegraphics[width=8cm]{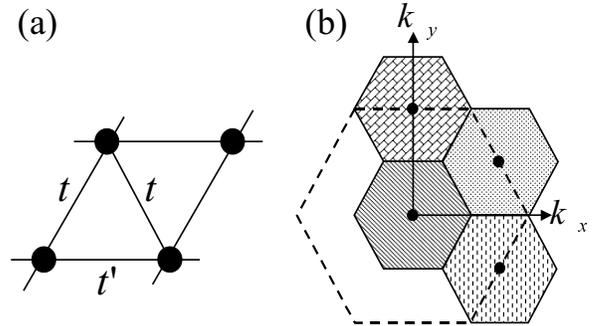}
\caption{(a) Schematic representation of triangular 
lattice with electron hoppings $t$ and $t'$. 
(b) Example of the coarse-graining cells in 
the BZ for the triangular lattice, 
where the cluster size $N_{c}$=4 and 
the dashed line denotes the first BZ. 
The dots represent the cluster momenta ${\bf K}$.
}
\label{fig:lattice}
\end{figure}

We start with the single-band Hubbard model, 
\begin{eqnarray}
H=
-t\sum_{\langle i,j \rangle \sigma}c^{\dag}_{i\sigma}c_{j\sigma}
-t'\sum_{\langle i,j' \rangle \sigma}c^{\dag}_{i\sigma}c_{j'\sigma} 
+U \sum_{i}n_{i\uparrow}n_{i\downarrow},
\end{eqnarray}
where $c_{i\sigma}(c^{\dag}_{i\sigma})$ is the annihilation 
(creation) operator of an electron with spin $\sigma$ at the 
$i$-th site, and $U$ represents the Coulomb repulsion.  
We introduce two kinds of hopping parameters $t$ and  $t'(\leq t)$
to study the effects of geometrical frustration systematically \cite{Kas01}
(FIG. \ref{fig:lattice} (a)).

In order to apply DCA to the frustrated lattice system, 
we first introduce the $N_c$ discrete cluster-momenta ${\bf K}$ 
\cite{Het98,Het00,Mai00,Mou01,Jar01}, 
which are defined as, 
\begin{eqnarray}
{\bf K}&=&n{\bf a}_{1}+m{\bf a}_{2}\\
{\bf a}_{1}&=&
\bigg(\frac{2\pi}{\sqrt{N_{c}}}, \frac{2\pi}{\sqrt{3N_{c}}}\bigg),\,\,
{\bf a}_{2}=\bigg(0, \frac{4\pi}{\sqrt{3N_{c}}}\bigg), 
\end{eqnarray}
where $n$ and $m$ denote the integral number 
and ${\bf K}$ should be within the first Brillouin zone (BZ). 
Next, we divide the original BZ
into the subregions specified by the 
cluster momenta (coarse-graining cells);
an example of $N_c=4$ is shown schematically 
in Fig. \ref{fig:lattice} (b). 
Therefore, in our treatment the coarse-grained Green 
function \cite{Het98,Het00,Mai00,Mou01,Jar01},
\begin{eqnarray}
\overline{G}_{\sigma}({\bf K},z)=\frac{1}{N'}\sum_{\tilde{\bf k}}
\frac{1}{z-\epsilon_{{\bf K+\tilde{k}}}-
\Sigma_{\sigma}({\bf K},z)},
\end{eqnarray}
is specified by the cluster momentum ${\bf K}$. 
Here, $N'=N/N_{c}$ where 
$N$ is the number of total lattice sites. 
The summation over ${\bf \tilde{k}}$ 
is taken within the coarse-graining cell. 

In order to obtain the coarse-grained Green function, we now map 
the Hubbard lattice model to the cluster Anderson model \cite{Mai00}. 
By solving this effective cluster problem, we can obtain the 
cluster self energy $\Sigma({\bf K},z)$, so that 
the coarse-grained Green function is determined self-consistently.
Then the one-particle spectral function and the 
total DOS are given by the standard formula,
\begin{eqnarray}
A_{\sigma}({\bf K}, \omega)
&=&-\frac{1}{\pi}{\rm Im}\overline{G}_{\sigma}({\bf K}, \omega)\\
\rho_{\sigma}(\omega)
&=&\frac{1}{N_{c}}\sum_{{\bf K}}
A_{\sigma}({\bf K}, \omega).
\end{eqnarray}

We numerically iterate the above procedure until the calculated 
quantities converge within desired accuracy. In the following 
discussions, we shall deal with a paramagnetic metallic phase.
We set $t=1$ as the unit of the energy for simplicity. 


In order to solve the cluster problem  mentioned 
above, we make use of NCA \cite{Mai00}, which is 
expected to provide reliable results in the temperature 
range we are now interested in. 
Since the dimension of the cluster Hamiltonian 
is $4^{N_{c}}$, so that the numerical 
calculation becomes much more difficult with the increase of 
the number of cluster momenta. We practically take  the 
cluster size, $N_{c}=4$  within NCA in this paper.
Note that we have  neglected off-diagonal terms in 
 the cluster resolvents, 
since those are known to be less
important for discussing the one-particle spectra \cite{Mai00}. 
We show our results calculated for $T=0.6$ below, since this 
temperature is reasonably low for our system to exhibit essential 
properties due to the electron correlations. 

\begin{figure}[htb]
\includegraphics[width=7cm]{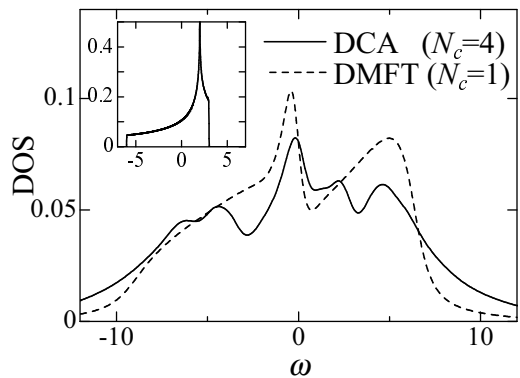}
\caption{Total DOS for the triangular lattice with 
DCA (solid line) and DMFT (dashed line) at half-filling, respectively. 
Here, the cluster size $N_{c}=4$. 
The energy 
is measured from the Fermi level ($\omega=0$). The parameters 
are chosen as $t'=t$, $U$=6.0 and  
the temperature $T$=0.6 ($t$ is taken 
as the unit of the energy). The inset shows DOS for the 
non-interacting tight-binding  model on the same lattice.
}
\label{fig:DOS}
\end{figure}

Let us first discuss the total DOS in the system with strong 
frustration. In FIG. \ref{fig:DOS}, we show the DOS for the 
triangular lattice at half-filling with
isotropic hopping, $t'=t$. 
For the triangular lattice, DOS calculated by DMFT has 
a many-body peak around the Fermi level ($\omega \sim 0$), 
which  implies  
the formation of heavy quasi-particles due to the Hubbard
interaction. 
We note here that the DCA recovers the DMFT for $N_{c}=1$. 
Even if short-range non-local correlations 
are taken into account by DCA, the many-body peak still persists 
although it becomes  slightly smaller than that of DMFT. 

\begin{figure}[htb]
\includegraphics[width=8cm]{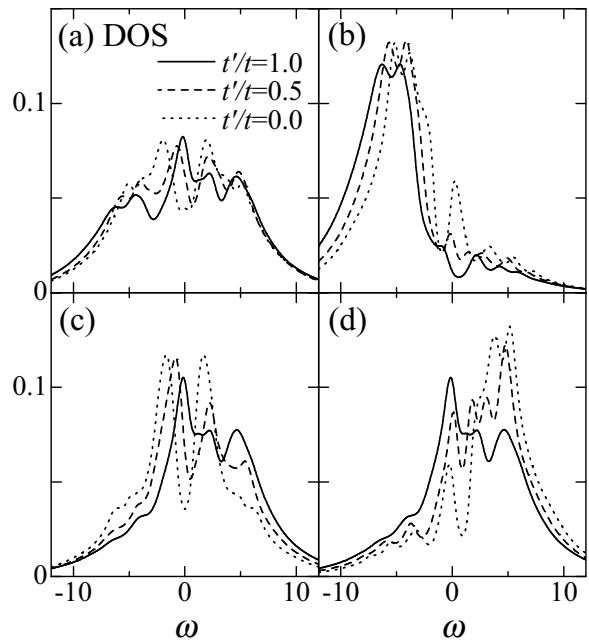}
\caption{(a) Total DOS $\rho (\omega)$  and (b)-(d) one-particle spectral
 functions $A({\bf K}, \omega)$ corresponding to  the 
${\bf K}=(0,0)$, $(\pi,\pi/\sqrt{3})$ (or $(\pi,-\pi/\sqrt{3})$) 
and $(0,2\pi/\sqrt{3})$ 
for various choices of the hopping amplitude: $t'=t$ (solid line), 
$t'=0.5t$ (dashed) and $t'=0.0$ (dotted). 
The other parameters are as in FIG. \ref{fig:DOS}. 
}
\label{fig:triangular}
\end{figure}

This Fermi-liquid like 
behavior is contrasted to the case of the square 
lattice with nearest neighbor hopping. In the latter case, it is known that
a pseudo-gap structure in DOS, which is 
developed by the Hubbard interaction, prevents the formation 
of quasi-particles \cite{Mai00} when the system is at half-filling
where short-range antiferromagnetic fluctuations 
are most relevant (see also  FIG. \ref{fig:triangular}). Our 
results in the triangular lattice shows
that antiferromagnetic correlations are strongly suppressed
in the frustrated lattices, resulting in the 
formation of heavy quasi-particles even at half filling.
Note that in both of these two lattices, the Mott insulator is 
stabilized as $U$ becomes large. 

To investigate how antiferromagnetic fluctuations are 
suppressed by 
geometrical frustration, we change $t'$ continuously, and 
observe what happens for  
DOS and the one-particle spectral functions on the triangular 
lattice. The results are shown in FIG. \ref{fig:triangular}. Note that 
the effect of frustration becomes less important
 with the decrease of $t'$.
As $t'$ decreases from $t$, the heavy quasi-particle band around 
the Fermi level is obscured, and then the pseudo-gap 
is developed after frustration 
is suppressed. From  the one-particle spectral function 
for each momentum,
shown in FIG. \ref{fig:triangular} (b)-(d), we can figure out
which contribution is most relevant to the quasi-particle
formation (or suppression) under strong frustration.
 Since the state with ${\bf K}=(0,0)$ is away from the Fermi level, 
the overall structure 
is not changed  by $t'$. However the spectrum for 
${\bf K}=(\pi,\pi/\sqrt{3})$, 
which is energetically 
degenerate with that for ${\bf K}=(\pi,-\pi/\sqrt{3})$, 
crosses the Fermi surface, being  easily 
affected by frustration. When $t'$ is equal to $t$, where 
frustration is very strong,  heavy quasi-particles are formed around 
the Fermi level. With decreasing frustration, the heavy quasi-particle 
band splits, and the pseudo-gap begins to develop explicitly 
around the Fermi level.
Therefore, the low energy physics in this system is mainly controlled by
the state with 
${\bf K}=(\pi,\pi/\sqrt{3})$ (or ${\bf K}=(\pi,-\pi/\sqrt{3})$). 

In order to further confirm the above 
statements, we have numerically 
estimated  the renormalization factor, 
which is defined as, 
\begin{eqnarray}
Z_{\bf K}=
\bigg(
1-\frac{\partial \Re \Sigma ({\bf K}, \omega )}{\partial \omega}
\bigg|_{\omega =0}\bigg)^{-1}.
\label{Z}
\end{eqnarray}
\begin{figure}[htb]
\includegraphics[width=8cm]{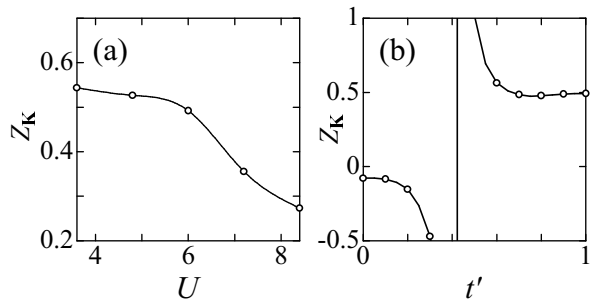}
\caption{Renormalization factor for the triangular lattice. 
(a) $U$-dependence ($t'=t$) and (b) $t'$-dependence ($U=6.0$). 
Here, ${\bf K}=(\pi,\pi/\sqrt{3})$.
The other parameters are as in FIG. \ref{fig:DOS}.
}
\label{fig:RZ}
\end{figure}

In FIG. \ref{fig:RZ} we show the results obtained for 
the momentum ${\bf K}=(\pi,\pi/\sqrt{3})$ close to
the Fermi level. It is seen from FIG. \ref{fig:RZ}(a) that 
the renormalization 
factor for the triangular lattice
 is reduced with the increase of $U$ for the fully frustrated 
case ($t'=t$).  As mentioned above, this indicates 
that the effective mass near the Fermi level is 
enhanced by $U$, forming well-defined heavy quasi-particles. 
When $t'/t$ is decreased from unity with $U$ being fixed, 
the effect of frustration becomes less important,
so that the system has a tendency to be an insulator with
strong antiferromagnetic fluctuations.
In this case,  we encounter an anomalous 
behavior in the renormalization factor calculated
numerically for the triangular lattice,
as seen from FIG.\ref{fig:RZ}.   Namely,
the renormalization factor increases as $t'$ decreases, and 
 diverges around $t'/t \sim 0.4$,  below which 
it takes negative values. This anomalous behavior  in 
$Z_{\bf K}$ is closely related to the formation of  
a dip-structure in the spectrum, which reflects the
enhancement of antiferromagnetic fluctuations.
The negative values of $Z_{\bf K}$ should be considered as an artifact of
fitting the numerical data with the formula (\ref{Z})
even for the case with a dip structure.

\begin{figure}[htb]
\includegraphics[width=7cm]{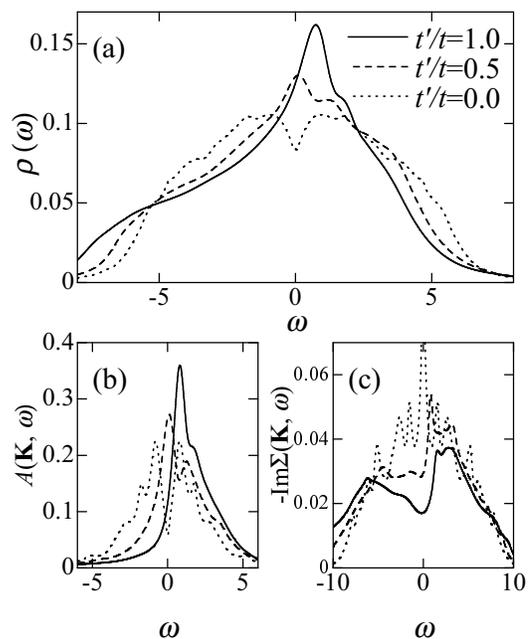}
\caption{(a) DOS $\rho (\omega)$, 
(b) one particle spectral function $A({\bf K},\omega)$ 
$({\bf K}=(\pi,\pi/\sqrt{3})$) and 
(c) imaginary part of the self-energy $\Sigma({\bf K},\omega)$ 
for $U=3.0$, $T=0.6$ at half-filling. 
The cluster size $N_{c}$ used for the calculation is 36.
}
\label{fig:DOS(FLEX)}
\end{figure}

So far, we have treated the effective cluster model 
by NCA, which may be an efficient approximation in the 
intermediate coupling regime.
Here, we take a weak-coupling approach 
 using FLEX to solve the effective cluster problem. 
Since FLEX is a perturbative method
with respect to the Coulomb interaction $U$, it may give results
complimentary to those of NCA.  We here confirm the conclusion of the 
NCA approach by investigating effects of frustration on
the triangular lattice in Fig.\ref{fig:lattice}(a).
Shown in Fig.\ref{fig:DOS(FLEX)} are the total DOS, the 
one-particle spectral function 
and the corresponding self-energy calculated for
 the triangular lattice at half-filling. 

In the case of $t'=0$, 
a pseudo-gap develops around the Fermi level in  the total DOS 
as well as in the one-particle spectral function 
as $U$ increases, which is consistent 
with the QMC results \cite{Mou01,Jar01}
and also with those obtained by NCA.  It is seen that 
the pseudo-gap disappears with the increase of $t'$, similarly to 
the results of NCA.  However, it is not clearly seen from this figure 
whether a heavy quasi-particle band is indeed formed, since
the bump structure in the DOS  above the Fermi 
level ($t'=0.5t$ and $1.0t$) is much affected by  
 the van-Hove singularity. Nevertheless, we can see
a tendency to the formation of Fermi quasi-particles  
by observing the imaginary part of the self-energy  in the 
large $t'$ regime. (Fig.\ref{fig:DOS(FLEX)}(c))
Namely, when $t'$ is absent, 
the imaginary part of the self-energy has a sharp peak 
structure around $\omega \sim 0$, which is quite different from
the Fermi-liquid behavior.  However, 
with increasing $t'$, its shape changes from the peak 
to a concave structure, which is similar to a
conventional Fermi liquid-like behavior.
We have also checked that qualitatively analogous behaviors 
 can be found in the system of different cluster size, $N_{c}=16$ and 64. 
Therefore, we confirm the conclusion of the
NCA approach that frustration assists 
the formation of Fermi quasi-particles by effectively 
suppressing the antiferromagnetic fluctuations.

We wish to mention here that our results are consistent 
with some experimental findings, e.g. the large specific-heat 
coefficient found in the pyrochlore compound, ${\rm Y(Sc)Mn_{2}}$ \cite{Bal96},
which is a typical example of geometrically frustrated metallic systems.

In summary, we have investigated the Hubbard model with geometrical 
frustration by employing the triangular lattice as a typical
example. Applying DCA combined with NCA and FLEX to the 
Hubbard model in a paramagnetic metallic phase close to the
Mott insulator, we have 
calculated  the one-particle spectral function.  
Based on the results obtained by both approaches,  
we have demonstrated how geometrical frustration suppresses 
antiferromagnetic correlations, and then assists the formation of a
 heavy quasi-particle band near the Mott insulating phase.

We acknowledge valuable discussions with A. Koga and A. Kawaguchi.
The work is partly supported by a Grant-in-Aid from the Ministry of 
Education, Science, Sports, and Culture. 
Parts of the numerical computations were done by the supercomputer center 
at Institute for Solid State Physics, The University of Tokyo.


\end{document}